\documentclass[preprint,aps]{revtex4-1}
\usepackage{graphicx}

\usepackage{soul}
\usepackage{epstopdf}
\usepackage{bm}
\usepackage{upgreek}
\usepackage{amsmath}
\usepackage{lipsum}
\usepackage{color,xcolor}
\begin{document}

%\title{Electronic Structure and Pseudogap like behavior in Double-Layer Nickelate La$_3$Ni$_2$O$_7$}

%\title{Orbital-Selective Nesting Driven Interlayer Antiferromagnetic Spin Density Wave in Trilayer Nickelate La$_4$Ni$_3$O$_{10}$}
%

%\title{Electronic Origin of Density Wave Orders in Trilayer Nickelate La$_4$Ni$_3$O$_{10}$}

\title{Electronic Origin of Density Wave Orders in a Trilayer Nickelate}

\author{Jiangang Yang$^{1,2,\sharp}$, Jun Zhan$^{1,2,\sharp}$, Taimin Miao$^{1,2,\sharp}$, Mengwu Huo$^{3,\sharp}$, Qichen Xu$^{1,2}$, Yinghao Li$^{1,2}$, Yuyang Xie$^{1,2}$, Bo Liang$^{1,2}$, Neng Cai$^{1,2}$, Hao Chen$^{1,2}$, Wenpei Zhu$^{1,2}$, Mingkai Xu$^{1,2}$, Shenjin Zhang$^{4}$, Fengfeng Zhang$^{4}$, Feng Yang$^{4}$, Zhimin Wang$^{4}$, Qinjun Peng$^{4}$, Hanqing Mao$^{1,2}$, Xintong Li$^{1,2}$, Zhihai Zhu$^{1,2}$, Guodong Liu$^{1,2}$, Zuyan Xu$^{4}$, Jiangping Hu$^{1,2}$, Xianxin Wu$^{5,*}$, Meng Wang$^{3,*}$, Lin Zhao$^{1,2,*}$ and X. J. Zhou$^{1,2,*}$
}

\affiliation{
\\$^{1}$Beijing National Laboratory for Condensed Matter Physics, Institute of Physics, Chinese Academy of Sciences, Beijing 100190, China.
\\$^{2}$School of Physical Sciences, University of Chinese Academy of Sciences, Beijing 100049, China.
\\$^{3}$Guangdong Provincial Key Laboratory of Magnetoelectric Physics and Devices, School of Physics, Sun Yat-Sen University, Guangzhou 510275, China.
\\$^{4}$Technical Institute of Physics and Chemistry, Chinese Academy of Sciences, Beijing 100190, China.
\\$^{5}$Institute of Theoretical Physics, Chinese Academy of Sciences, Beijing 100190, China. 
\\$^{\sharp}$These people contribute equally to the present work.
\\$^{*}$Corresponding authors: XJZhou@iphy.ac.cn, LZhao@iphy.ac.cn, wangmeng5@mail.sysu.edu.cn, xxwu@itp.ac.cn
}

\date{\today}
%\pacs{}

\maketitle

\newpage

{\bf 
The discovery of superconductivity in Ruddlesden-Popper nickelates has established a new frontier in the study of high-temperature superconductors. However, the underlying pairing mechanism and its relationship to the material’s electronic and magnetic ground states remain elusive. Since unconventional superconductivity often emerges from a complex interplay of magnetic correlations, elucidating the magnetic ground state of the nickelates at ambient pressure is crucial for understanding the emergence of superconductivity under high pressure. Here, we combine high-resolution angle-resolved photoemission spectroscopy with tight-binding model simulation to investigate the electronic structure of the representative trilayer Ruddlesden-Popper nickelate La$_4$Ni$_3$O$_{10}$. We provide the first experimental evidence of band splitting induced by interlayer coupling and further resolve the momentum-dependent density wave gap structures along all the Fermi surfaces. Our findings identify the mirror-selective Fermi surface nesting as the origin of the interlayer antiferromagnetic spin density wave and demonstrate the dominant role of $\text{Ni-}3d_{z^2}$ orbitals in the low-energy physics of $\text{La}_4\text{Ni}_3\text{O}_{10}$. These results provide a fundamental framework for understanding the magnetic interactions and high-temperature superconductivity mechanism in the Ruddlesden-Popper nickelate family.
}

~\\
\noindent{\bf\large Introduction}\\

The recent discovery of high-temperature superconductivity in Ruddlesden-Popper (RP) nickelates has established a novel platform for investigating unconventional superconductivity\cite{MWang2023HLSun,HQYuan2024YNZhang_NP,JGCheng2024NNWang_Nature,JZhao2024YHZhu,ZYChen2025GDZhou_Nature,HYHwang2025_Nature,ZhangJJ2025Nature_327}. Research into these materials not only offers critical insights into the pairing mechanisms of cuprates and iron-based superconductors but also promises to unveil an entirely new paradigm that goes beyond these established frameworks\cite{JZaanen2015BKeimer,FernandesR2022Review,KeimerB2025Review,WangM2024Review_CPL,ChenXH2025review_NSR}.  Superconductivity in nickelates was initially discovered in the infinite-layer system $R$NiO$_2$ ($R$ = rare-earth element)\cite{HYHwang2019DFLi,DFLi2021}, whose pairing mechanism has been interpreted through an analogue to the in-plane physics of cuprates, primarily governed by the correlation effects of 3$d_{x^2-y^2}$ orbitals within the quasi-two-dimensional Ni-O planes\cite{NormanMR2020PRX_Ni112,LechermannF2020PRB_Ni112,SakakibaraH2020PRL_Ni112,WuXX2020PRB_Ni112,ZhangGM2020PRB_Ni112}. However, the recent discovery of high-Tc superconductivity in RP nickelates under pressure\cite{MWang2023HLSun,HQYuan2024YNZhang_NP,JGCheng2024NNWang_Nature,JZhao2024YHZhu,YPQi2025MXZhang_PRX,XHChen2025MZShi_NP,ZhangJJ2025Nature_327} or interfacial structure\cite{ZYChen2025GDZhou_Nature,HYHwang2025_Nature}, most notably in the bilayer R$_3$Ni$_2$O$_7$\cite{MWang2023HLSun,HQYuan2024YNZhang_NP,JGCheng2024NNWang_Nature,ZhangJJ2025Nature_327,ZYChen2025GDZhou_Nature,HYHwang2025_Nature}, the trilayer R$_4$Ni$_3$O$_{10}$\cite{JZhao2024YHZhu,YPQi2025MXZhang_PRX} and even more complex hybridized intergrowth phases\cite{XHChen2025MZShi_NP,WangM2025HuangC_arXiv}, has brought a new dimension to the field. In these multi-layer systems, the emergence of the 3$d_{z^2}$ orbital degree of freedom and the predominant role of interlayer coupling open a new window into the exploration of unconventional superconductivity.

In strongly correlated systems, unconventional superconductivity is intimately entangled with competing quantum orders such as density wave states and magnetic order\cite{JZaanen2015BKeimer,FernandesR2022Review,KeimerB2025Review,WangM2024Review_CPL,ChenXH2025review_NSR}. It is generally believed that suppressing these orders toward a quantum critical point via doping or strain enhances quantum fluctuations, thereby promoting superconducting pairing. Elucidating the mechanisms for the formation of these electronic orders is essential to unraveling the microscopic framework of unconventional superconductivity. In RP phase nickelates, substantial evidence points to existence of competing density wave orders while pressure-mediated suppression of these ordered states leads to the emergence of superconductivity\cite{MitchellJF2020ZhangJJ_NC,WangM2024HuangX_CPL,FengDL2024ChenXY_NC,MukudaH2024_JPSJ,ShuL2024PRL_miuSR,ZhuZH2025CP,WangNL2025XuSX_PRB,YanYJ2025LiMZ_PRB,GuoHJ2025CaoYT,ZhaoJ2025ZhangEK_PRX,GuguchiaZ2025KhasanovR,YangLX2025LiYD_SB}. Moreover, the presence of these density-wave orders at ambient pressure is found to be intimately tied to the emergence of superconductivity under high pressure\cite{ChenXH2025ShiMZ_NC,WangNL2025XuSX_NC,ChenXH2025arXiv_327}.

As a prototypical member of the trilayer RP nickelates, La$_4$Ni$_3$O$_{10}$ has recently been discovered to host bulk superconductivity with a critical temperature  exceeding 30\,K under high pressure\cite{JZhao2024YHZhu,YPQi2025MXZhang_PRX}, establishing this material as an ideal system for exploring the interplay between multi-orbital correlations, interlayer coupling, and unconventional pairing. Compared to its bilayer counterparts La$_3$Ni$_2$O$_{7}$, the superior crystal quality and relative ease of synthesis make La$_4$Ni$_3$O$_{10}$ more favorable for high-precision spectroscopic investigations. At ambient pressure, transport and magnetic susceptibility measurements reveal a prominent metal-to-metal transition at $T^* \approx 140$\,K, accompanied by a transition in magnetic structure\cite{MitchellJF2020ZhangJJ_NC}. This transition is progressively suppressed by external pressure, eventually giving way to the superconducting state\cite{JZhao2024YHZhu,YPQi2025MXZhang_PRX}. X-ray and neutron scattering studies have demonstrated that this transition originates from the formation of entangled spin- and charge-density wave orders\cite{MitchellJF2020ZhangJJ_NC}. Although neutron diffraction and muon-spin rotation/relaxation (µSR) experiments have elucidated the interlayer antiferromagnetic nature of spin density wave (SDW)\cite{MitchellJF2020ZhangJJ_NC,GuguchiaZ2025arXiv_miuSR}, the underlying mechanism and specific contributions from distinct orbital degrees of freedom remain elusive. Specifically, whether the SDW is driven by itinerant Fermi surface nesting or localized magnetic interaction, and how the structure symmetry and orbital-selective physics constrains these interactions, remains unclear. Although previous angle-resolved photoemission spectroscopy (ARPES) studies have observed energy gap openings associated with the density wave transition, the reported results exhibit significant discrepancies and remain under debate\cite{DSDessau2017HXLi, YangLX2024DuX_arXiv}. Critically, a unified, momentum-resolved determination of the gap evolution across the entire Fermi surface is still missing which is essential for elucidating the formation mechanism of this competing order and for uncovering the decisive roles played by orbital degrees of freedom and crystal symmetry.

In this work, by using high-resolution laser-based and synchrotron-based ARPES combined with tight-binding model simulation, we provide a complete picture of the low-energy electronic structure and density wave induced gap opening in La$_4$Ni$_3$O$_{10}$. For the first time, we resolve the complete four Fermi surface sheets in La$_4$Ni$_3$O$_{10}$, including the two Fermi surface sheets induced by the interlayer coupling. Density wave induced gap opening is observed below $T^* \approx 140$\,K which exhibits distinct momentum dependence on different Fermi surface sheets. Crucially, we identify that a phase-locked interlayer antiferromagnetic SDW order is driven by Fermi surface scattering between bands with opposite mirror parity. By correlating the gap anisotropy with the orbital-projected Fermi surface topology, we demonstrate that $d_{z^2}$-mediated interlayer interactions are the primary driver of this spin density wave state. These findings indicate that the suppression of such SDW orders under pressure likely releases the interlayer spin fluctuations which is conducive to superconducting pairing in the RP nickelate family.

~\\
\noindent {\bf\large Results}\\

The crystal structure of La$_4$Ni$_3$O$_{10}$ comprises triple-layer units of corner-sharing NiO$_6$ octahedra, which are stacked along the $c$-axis and separated by rock-salt-type La-O layers. At ambient pressure, the structure deviates from ideal tetragonal symmetry due to the rotation and tilting of the NiO$_6$ octahedra, resulting in a subtle symmetry breaking that reduces the space group to $P2_1/a$ (monoclinic)\cite{MitchelJ2020ZhangJJ_PRM}. Under high pressure, the system transitions into a tetragonal phase with the $I4/mmm$ space group, where superconductivity emerges\cite{JZhao2024YHZhu,WangM2024SCPMA_Ni4310}. This lattice distortion from the tetragonal to the monoclinic phase leads to a Brillouin zone (BZ) folding, accompanied by the replication of electronic bands. To facilitate our theoretical modeling and maintain consistency with discussions regarding the high-pressure physics, all experimental data in this study are mapped onto the large Brillouin zone defined by the tetragonal symmetry.

To directly probe the electronic structure of La$_4$Ni$_3$O$_{10}$, we performed high-resolution ARPES using both laser and synchrotron light sources. Figures 1a–1d show the Fermi surface mappings and constant energy contours at 30\,meV binding energy under different polarization geometries measured by laser ARPES with the photon energy of 6.994\,eV. Strong photoemission matrix element effects are observed due to the multi-orbital nature of La$_4$Ni$_3$O$_{10}$. Fig. 1e shows the Fermi surface mapping measured by synchrotron-based ARPES with a photon energy of 85\,eV. From these Fermi surface mappings and constant energy contours (Figs. 1a-1e), combined with the related band structure analysis (Fig. 2), we resolve a complete Fermi surface topology of La$_4$Ni$_3$O$_{10}$, as shown in Fig. 1f. These include four main Fermi surface sheets that consist of an electron-like pocket $\alpha$ centered at the $\Gamma$ point and three hole-like pockets ($\beta$, $\beta'$ and $\gamma$) centered at the M point. Notably, the $\beta'$ sheet is identified for the first time which originates from the band splitting induced by the interlayer coupling. Furthermore, we observe several folded bands arising from crystal reconstruction or density-wave order. Specifically, the $\sqrt{2} \times \sqrt{2}$ reconstruction gives rise to $\alpha_f$, $\beta_f$, $\beta'_f$ and $\gamma_f$ sheets as plotted in Fig. 1f. The $\beta_f$ sheet, folded from the $\beta$ Fermi surface, is particularly strong and clear in both polarization geometries (Fig. 1a and 1c). The $\delta$ feature is similar to that observed in  La$_3$Ni$_2$O$_{7}$\cite{YangJG2024_NC,Damascelli2025arXiv} which is attributed to band folding driven by the spin density wave (SDW) as we will discuss later.

To gain deeper insights into the underlying electronic structure, we constructed a trilayer two-orbital tight-binding model, as schematically illustrated in Fig. 1g. Our model considers a trilayer lattice that incorporates both Ni $3d_{x^2-y^2}$ and the $3d_{z^2}$ orbitals at each site. The model includes both intra-layer and inter-layer hoppings between these on-site orbitals (see Supplementary Information for details).  We optimized the tight-binding parameters to capture the measured Fermi surface topology (Fig. 1f) and the band structures (Fig. 2), with the parameters summarized in the Supplementary Information. Fig. 1h shows the simulated Fermi surface from our tight-binding model which is well matched with our measured Fermi surface topology. This model provides a foundation for further analysis of the band structures, many-body effects and the density wave instability in La$_4$Ni$_3$O$_{10}$.

Figure 2 shows the detailed momentum-dependent band structures of La$_4$Ni$_3$O$_{10}$. Fig. 2a presents the band structures measured at 20\,K along high-symmetry directions by using synchrotron-based ARPES with 85\,eV photon energy. Here the main $\alpha$, $\beta$ and $\gamma$ bands are clearly observed as indicated by the colored arrows. The simulated band structures in Fig. 2b well describe the measured main $\alpha$, $\beta$ and $\gamma$ bands although the $\beta'$ band is not well resolved in the measurements. To resolve fine band structures of La$_4$Ni$_3$O$_{10}$, we took high-resolution laser-based ARPES measurements under different polarization geometries. Under the s polarization geometry, the spectral weight of the main bands is strong near the $\overline{\Gamma}-\overline{M}$ diagonal region (Fig. 2c). Under the p polarization geometry, the spectral weight of the main  $\beta$ and $\beta'$ bands is strong away from the $\overline{\Gamma}-\overline{M}$ diagonal region (Fig. 2d). A detailed momentum dependence of the band structures around the $\overline{\Gamma}-\overline{M}$ diagonal region is presented in Fig. 2e obtained from the $s$ polarization measurements in Fig. 2c. To better resolve the observed bands, Fig. 2f shows the corresponding momentum distribution curves (MDC) second-derivative images. Here the main $\alpha$, $\beta$ and $\beta'$ bands are clearly resolved as marked by the black, red and green arrows, respectively. The corresponding simulated band structures in Fig. 2g are consistent with the measured results. Turning to the $p$ polarization measurements (Fig. 2d), detailed momentum-dependent band structures away from the $\overline{\Gamma}-\overline{M}$ diagonal region are shown in Fig. 2h and the corresponding MDC second-derivative images are presented in Fig. 2i. Here the main $\beta$ and $\beta'$ bands are clearly resolved as marked by the red and green arrows. The $\delta$ band is also clearly observed (as marked by purple arrows in Fig. 2h) which is attributed to the SDW induced band folding. The simulated band structures in Fig. 2j well capture the measured band structures in Fig. 2h and 2i. 

Notably, the main $\alpha$, $\beta'$ and $\gamma$ bands exhibit a clear gap opening at low temperature for some momentum cuts, as seen in Fig. 2a and Fig. 2e. To obtain a complete mapping of the gap structure, we systematically investigated the momentum dependence of the gap along all the observed Fermi surface sheets. Fig. 3a–3e presents the symmetrized EDCs measured at 20\,K along the $\alpha$, $\beta$, $\beta_f$, $\beta'$ and $\gamma$ Fermi surface sheets, respectively (the corresponding original EDCs are shown in  Fig. S1 in Supplementary Information). The gap size can be determined from the peak position in the symmetrized EDCs, as indicated by the ticks in Fig. 3a-3e. The extracted gap distribution along these Fermi surface sheets is plotted in Fig. 3f which shows the gap magnitude $\Delta$ as a function of the Fermi surface angle $\theta$ (defined in top panels of Figs. 3a–3e). The gap distribution along different Fermi surface sheets is schematically shown in the three dimensional plot in Fig. 3g. The four main Fermi surface sheets exhibit distinct momentum dependence of the energy gap. The gap opening along the $\alpha$ Fermi surface is quite anisotropic with its gap minimum along the $\overline{\Gamma}-\overline{M}$ diagonal direction and the gap maximum along the $\overline{\Gamma}-\overline{X}$ direction reaching a gap size of $\sim$33\,meV. In contrast, no discernible gap opening is observed on the $\beta$ Fermi surface or its structural folding-induced $\beta_f$ sheet. Interestingly, for the $\beta'$ Fermi surface, an anisotropic energy gap opens along the central part of the Fermi surface near the $\overline{\Gamma}-\overline{M}$ diagonal region but the gap vanishes away from the diagonal region extending to the $\overline{X}$ point. Furthermore, the gap opening along the $\gamma$ Fermi surface is nearly isotropic with the gap size of $\sim$33\,meV. This comprehensive, momentum-resolved gap landscape provides crucial insights into the complex low-energy physics and the origin of the density wave formations in La$_4$Ni$_3$O$_{10}$.

It is found that La$_4$Ni$_3$O$_{10}$ exhibits a phase transition around 140\,K which is believed to be associated with the coexistence of the SDW and charge density wave (CDW) transitions\cite{MitchellJF2020ZhangJJ_NC,GuguchiaZ2025arXiv_miuSR}.  To investigate whether the gap opening we observed at low temperature is related to such transitions, we further carried out systematic temperature dependent measurements on the $\alpha$ and $\gamma$ bands. Fig. 4a shows the band structures of the $\alpha$ band measured at different temperatures along the $\overline{\Gamma}-\overline{X}$ direction where the gap opening is maximal. The $\alpha$ band shows a clear spectral weight suppression near the Fermi level at low temperatures, signaling the opening of an energy gap. To determine the gap quantitatively, we extracted EDCs at the Fermi momentum ($k_\mathrm{F}$) of the $\alpha$ band at different temperatures (Fig. 4c). To extract the gap size, these EDCs are symmetrized to remove the Fermi-Dirac distribution function and the gap size is obtained from the peak position of the symmetrized EDCs as marked by ticks in Fig. 4d. The temperature dependence of the gap size for the $\alpha$ band is plotted in Fig. 4h. The gap size of the $\alpha$ band decreases with the increase of the temperature and approaches nearly zero at $\sim$140\,K which is consitant with the temperature of the SDW/CDW transitions as determined from the resistivity (Fig. 4i) and magnetic susceptibility (Fig. 4j) measurements. Fig. 4e shows the band structures of the $\gamma$ band measured at different temperatures along the $\overline{M}-\overline{X}-\overline{M}$ direction. The $\gamma$ band shows a clear spectral weight suppression near the Fermi level at low temperatures, signaling the opening of an energy gap. The EDCs at the Fermi momentum  of the $\gamma$ band at different temperatures are shown in Fig. 4f, and the corresponding symmetrized EDCs are presented in Fig. 4g. The extracted gap size of the $\gamma$ band at different temperatures is also included in Fig. 4h. Within experimental uncertainty, the gap opening of the $\gamma$ band exhibits similar behavior to that of the $\alpha$ band, in terms of both the gap size and the temperature dependence. These results clearly indicate that the gap opening of the $\alpha$ and $\gamma$ bands is intimately related with the SDW/CDW transitions.

As shown in Fig. 1 and Fig. 2, the low-energy electronic structure of La$_4$Ni$_3$O$_{10}$ is well captured by our trilayer two-orbital tight-binding model (Note 1 in Supplementary Information). We further use this model to elucidate the roles of interlayer coupling, multiorbital degrees of freedom and crystalline symmetry in the low energy physics of $\text{La}_4\text{Ni}_3\text{O}_{10}$. Fig. 5a shows the calculated band structures along the high-symmetry directions. Driven by the strong interlayer coupling of the $d_{z^2}$ orbitals, these bands are classified into three distinct groups: bonding, anti-bonding and non-bonding states. Specifically, the $\alpha$ and $\gamma$ bands exhibit bonding character, while the $\beta$ and $\beta'$ bands are identified as anti-bonding and non-bonding states, respectively. Under mirror symmetry with respect to the inner Ni-O plane, the bonding and anti-bonding bands ($\alpha$, $\beta$ and $\gamma$) possess even (+) parity, whereas the non-bonding band ($\beta'$) is characterized by odd (-) parity. Fig. 5b displays the calculated Fermi surface topology, with the color representing the projection of orbital degrees of freedom. Near the $\Gamma$-M diagonal region, the $\alpha$, $\beta$ and $\beta'$ Fermi surfaces are predominantly composed of $d_{x^2-y^2}$ orbitals which gradually show prominent $d_{z^2}$ orbital admixture when moving away from this region. In contrast, the $\gamma$ Fermi surface is dominated by $d_{z^2}$ orbital character. 

From the calculated Fermi surface topology in Fig. 5b, we can identify some potential inter-band and intra-band nestings. The electron-like $\alpha$ Fermi surface shows good nesting with the hole-like $\beta$, $\beta'$ and $\gamma$ Fermi surfaces as indicated by the arrowed lines $\mathbf{Q}_1$ and $\mathbf{Q}_2$. In addition, the $\beta$ and $\beta'$ Fermi surfaces exhibit good intra-band nesting near the $X$ region as indicated by the arrowed lines $\mathbf{Q}_3$ and $\mathbf{Q}_4$. To substantiate these observations, we calculated the multiorbital susceptibility based on our tight-binding model (see Note 2 in Supplementary Information). Due to the approximate mirror symmetry, the susceptibility can be decoupled into mirror-odd ($\chi_{\text{odd}}$) and mirror-even ($\chi_{\text{even}}$) channels as shown in Fig. 5c and 5d. The mirror-odd susceptibility accounts for the scattering between Fermi surfaces with opposite mirror parity. It exhibits a pronounced peak at the wavevector $\mathbf{Q}_1 \approx (0.62, 0.62)\pi/a$ which primarily originates from the nesting between the even-parity $\alpha$ Fermi surface and the odd-parity $\beta'$ Fermi surface (Fig. 5b). Notably, this wavevector is in a good agreement with the SDW wavevector reported in neutron scattering measurements\cite{MitchellJF2020ZhangJJ_NC}. Moreover, the resulting real-space SDW pattern driven by this nesting (Fig. 5e) is mirror-odd which yields a negligible moment on the inner Ni-O layer and mediates antiferromagnetic correlations between the outer Ni-O layers (Fig. 5f), also consistent with neutron scattering measurements\cite{MitchellJF2020ZhangJJ_NC}. These results suggest that the energy gaps we observed on the $\alpha$ and $\beta'$ Fermi surfaces (Fig. 4) can be attributed to the itinerant SDW instability driven by mirror-selective scattering. The mirror-even channel susceptibility in Fig. 5d accounts for the scattering between Fermi surfaces with the same parity. Several scattering vectors emerge in this channel, including $\mathbf{Q}_2 \approx (1, 1)\pi/a$, $\mathbf{Q}_3 \approx (0.36, 0)\pi/a$ and $\mathbf{Q}_4 \approx (0.45, 0)\pi/a$. The scattering vector $\mathbf{Q}_2$ arises from the nesting between the $\alpha$ and $\gamma$ Fermi surfaces, while $\mathbf{Q}_3$ and $\mathbf{Q}_4$ are attributed to $\beta$-$\beta$ and $\beta'$-$\beta'$ intra-band nesting, respectively. For $\mathbf{Q}_3$ and $\mathbf{Q}_4$, no gap opening is observed on the related Fermi surface sections. However, for $\mathbf{Q}_2$, we propose that the nesting is likely responsible for the gap opening observed on the $\gamma$ band. While there are currently no experimental reports of a density wave at this specific wavevector, we note that $\mathbf{Q}_2$ coincides with the $\sqrt{2} \times \sqrt{2}$ Brillouin zone folding induced by structural reconstruction. Consequently, the density wave signal might be overlapping with the structural Bragg peaks in scattering experiments. Furthermore, this nesting-driven instability may further couple with the lattice degree of freedom and enhance the $\sqrt{2} \times \sqrt{2}$ lattice reconstruction at low temperatures.

The role of electronic correlations was further examined using the many-body functional renormalization group (FRG) approach. As the FRG energy scale decreases, the mirror-odd SDW channel shows the most prominent divergence (Fig. S2a), with the effective scattering strength peaking at $\mathbf{Q}_1 \approx (0.62, 0.62)\pi/a$ (Fig. S2b). The corresponding leading eigenmode (Fig. S2c) reveals a mirror-odd interlayer antiferromagnetic configuration. This consistent correlation-driven picture reinforces the mirror-selective scattering origin of the itinerant SDW order in $\text{La}_4\text{Ni}_3\text{O}_{10}$.

The presence of SDW order is further substantiated by the band folding observed in our ARPES measurements (Fig. 1 and Fig. 2), which provides a natural explanation for the $\delta$ Fermi surface sheet. To elucidate this, we performed layer-resolved weight projections of the Fermi surface, as shown in Fig. S3 which reveals that the $\beta'$ band possesses a strong outer-layer character in regions away from the $\Gamma$-M diagonals. Since the $\mathbf{Q}_1$ SDW order is primarily localized on the outer Ni-O layers, it is expected to induce a pronounced band reconstruction specifically within these outer-layer-dominated states. As illustrated in Fig. S4, the $\beta'$ band, when shifted by the SDW wavevector $\mathbf{Q}_1$, produces a folded electronic structure. It well matches the geometry of the $\delta$ Fermi surface sheet observed from ARPES measurements which confirm the SDW-induced folding origin of the $\delta$ band.

By further incorporating the orbital degree of freedom, we observe a strong correlation between the momentum-dependent magnitude of the energy gaps and the distribution of the $d_{z^2}$ orbital weight along the Fermi surface. The gap anisotropy on the $\alpha$ sheet is closely tied to its $d_{z^2}$ orbital composition, with the maximum gap appearing along the $d_{z^2}$-rich $\Gamma-X$ direction and the minimum residing along the $\Gamma-M$ direction where the $d_{z^2}$ weight is negligible. This correlation indicates that the nesting-driven Fermi surface instabilities in $\text{La}_4\text{Ni}_3\text{O}_{10}$ are primarily mediated by the Ni-$d_{z^2}$ orbitals. This orbital-selective nature of the electronic instability suggests that the interlayer coupling, facilitated by the $d_{z^2}$ and apical oxygen orbitals, is the decisive factor in establishing the long-range ordered states in this trilayer system. While the $d_{x^2-y^2}$ electrons maintain the metallic framework of the Ni-O planes, the $d_{z^2}$ electrons act as the primary glue that couples the layers and drives the system toward SDW instability via scattering in the mirror-odd channel, which can also induce a concomitant CDW.

~\\
~\\
\noindent{\bf\large{Discussion}}

The discovery of high-temperature superconductivity in the RP nickelates presents a unique paradigm that potentially transcends the established frameworks of both cuprates and iron-based superconductors. In cuprates, the physics is predominantly governed by the doping of a Mott insulator, where an antiferromagnetic order driven by strong localized single-orbital $d_{x^2-y^2}$ physics is suppressed to release the spin fluctuations necessary for pairing\cite{JZaanen2015BKeimer}. This creates a manifest competition between the long-range antiferromagnetic state and superconductivity, confined largely to the two-dimensional CuO$_2$ planes. Conversely, iron-based superconductors are characterized by itinerant antiferromagnetic spin fluctuations driven by Fermi surface nesting within a multi-orbital landscape\cite{FernandesR2022Review}. While interlayer coupling has been noted to enhance the transition temperature\cite{XJZhou2023LuoXY}, the fundamental pairing glue in both cuprates and pnictides is widely regarded as arising from in-plane magnetic interactions, effectively maintaining a 2D-dominant pairing mechanism\cite{QMSi2016Review_NatureRP,SchmalianJ2025Review}. However, the RP nickelates present a unique structural and electronic landscape: the superconducting units feature strongly coupled layers and the low-energy physics is governed by both in-plane $d_{x^2-y^2}$  and out-of-plane $d_{z^2}$ orbitals. While the superconducting mechanism is still under debate regarding whether the pairing originates primarily from the in-plane interactions or interlayer couplings \cite{WangQH2023YangQG_PRB,ZhangGM2023CPL,HuJP2025GuYH_PRB,YaoDX2024LuoZH_NPJ,YangF2023LiuYB_PRB,WangQH2024YangQG_PRB_4310,KurokiK2024PRB,YangF2025ZhangM_PRB,ZhouT2024_PRB,YangF2024ZhangM_PRB,WuCJ2025LuC_PRB,LeonovIV2024PRB_4310,XiangT2024FanZ_PRB,ZhangFC2024JiangK_CPL,ChenHH2025_NC},  $d_{z^2}$-derived interlayer coupling is widely believed to be essential\cite{MWang2023HLSun,ZhangGM2023CPL,WangQH2023YangQG_PRB,HuJP2025GuYH_PRB}.

Our ARPES findings provide critical insights into the interplay between the magnetic order and superconductivity in the trilayer RP nickelate La$_4$Ni$_3$O$_{10}$. We unveil a mirror-selective, nesting-driven SDW supported by direct spectroscopic evidence of Fermi surface nesting. Furthermore, our unbiased many-body calculations confirm that electronic correlations enhance the mirror-odd nesting, inducing a mirror-odd SDW instability. This suggests that the scattering between states of opposite mirror parity underpins SDW formation \cite{WuXX2025_arXiv_327}. By further comparing the momentum dependence of the gap structure with the orbital-resolved spectral weight, we demonstrate that the $\text{Ni-}d_{z^2}$ orbital is central to this interlayer magnetic order. These results provide compelling evidence that $d_{z^2}$-mediated interlayer antiferromagnetic interactions dominate the low-energy physics of $\text{La}_4\text{Ni}_3\text{O}_{10}$. Consequently, this offers a potential mechanism for superconductivity: when high pressure suppresses the static magnetic order, the resulting strong interlayer spin fluctuations may facilitate interlayer superconducting pairing. Given the structural similarities across the RP nickelates ($n > 1$), it is plausible to extend this physical framework to the entire family to understand the universally observed superconductivity therein. In conclusion, our results offer powerful insights into the complex magnetic interactions and potential pairing mechanisms in RP nickelates, suggesting that the RP family represents a distinct class of materials where unconventional superconductivity can be realized through the strategic tuning of 3D interlayer magnetic coupling.

~\\

\noindent{\bf\large Methods}

\noindent{\bf Growth of single crystals.} High-quality single crystals of trilayer nickelate La$_4$Ni$_3$O$_{10}$ were grown by using a high-pressure floating zone method\cite{WangM2024SCPMA_Ni4310}.  The crystals were characterized by X-ray diffraction analysis. Transport and magnetic susceptibility measurements were performed on the crystals. \\

\noindent{\bf ARPES measurements.} Synchrotron-based ARPES measurements were performed at the beamline BL03U of the Shanghai Synchrotron Radiation Facility (SSRF) with a hemispherical electron energy analyzer DA30L (Scienta-Omicron). The energy resolution was set at 10$\sim$15\,meV. High-resolution ARPES measurements were also performed using a lab-based ARPES system equipped with the 6.994\,eV vacuum-ultra-violet (VUV) laser and a hemispherical electron energy analyzer DA30L (Scienta-Omicron)\cite{XJZhou2008GDLiu,XJZhou2018}. The energy resolution was set at 2\,meV and the angular resolution was 0.2 degree. The momentum coverage is increased by applying bias voltage on the sample during the ARPES measurements\cite{MiaoTM2025arXiv}. All the samples were cleaved {\it in situ} at a low temperature of 20\,K and measured in ultrahigh vacuum with a base pressure better than 5 x 10$^{-11}$\,mbar. The Fermi level is referenced by measuring on clean polycrystalline gold that is electrically connected to the sample.

~\\
\noindent {\bf\large Data availability}\\
All data are processed by using Igor Pro 8.02 software. All data needed to evaluate the
conclusions in the paper are available within the article. All raw data generated during the current study are available from the
corresponding author upon request.

~\\
\noindent {\bf\large Code availability}\\
The codes used for the Tight-binding calculations in this study are available from the corresponding authors upon request.

~\\
\noindent {\bf\large References}

\bibliographystyle{unsrt}

\begin{thebibliography}{10}

\bibitem{MWang2023HLSun}
Sun, H. et al.
\newblock Signatures of superconductivity near 80\,K in a nickelate under high pressure.
\newblock {\em Nature} 621, 493--498 (2023).

\bibitem{HQYuan2024YNZhang_NP}
Zhang, Y. et al.
\newblock High-temperature superconductivity with zero resistance and strange-metal behaviour in $\text{La}_3\text{Ni}_2\text{O}_{7-\delta}$.
\newblock {\em Nature Physics} 20, 1269--1273 (2024).

\bibitem{JGCheng2024NNWang_Nature}
Wang, N. et al.
\newblock Bulk high-temperature superconductivity in pressurized tetragonal $\text{La}_2\text{Pr}\text{Ni}_2\text{O}_7$.
\newblock {\em Nature} 634, 579--584 (2024).

\bibitem{JZhao2024YHZhu}
Zhu, Y. et al.
\newblock Superconductivity in pressurized trilayer $\text{La}_4\text{Ni}_3\text{O}_{10-\delta}$ single crystals.
\newblock {\em Nature} 631, 531--536 (2024).

\bibitem{ZYChen2025GDZhou_Nature}
Zhou, G. et al.
\newblock Ambient-pressure superconductivity onset above 40 K in $(\text{La}, \text{Pr})_3\text{Ni}_2\text{O}_7$ films.
\newblock {\em Nature} 640, 641--646 (2025).

\bibitem{HYHwang2025_Nature}
Ko, E. K. et~al.
\newblock Signatures of ambient pressure superconductivity in thin film $\text{La}_3\text{Ni}_2\text{O}_{7}$.
\newblock {\em Nature} 638, 935--940 (2025).

\bibitem{ZhangJJ2025Nature_327}
Li, F. et~al. 
\newblock Bulk superconductivity up to 96 K in pressurized nickelate single crystals.
\newblock {\em Nature} 649, 871--878 (2026).

\bibitem{JZaanen2015BKeimer}
Keimer, B. et~al.
\newblock {From quantum matter to high-temperature superconductivity in copper oxides}.
\newblock {\em Nature} 518, 179--186 (2015).

\bibitem{FernandesR2022Review}
Fernandes, R.~M. et~al. 
\newblock Iron pnictides and chalcogenides: a new paradigm for superconductivity.
\newblock {\em Nature} 601, 35–44 (2022).

\bibitem{KeimerB2025Review}
Puphal, P. et~al. 
\newblock Superconductivity in infinite-layer and Ruddlesden–Popper nickelates.
\newblock {\em Nature Reviews Physics} (2025). https://doi.org/10.1038/s42254-025-00898-2

\bibitem{WangM2024Review_CPL}
Wang, M. et~al. 
\newblock Normal and superconducting properties of $\text{La}_3\text{Ni}_2\text{O}_{7}$.
\newblock {\em Chinese Physics Letters} 41, 077402 (2024).

\bibitem{ChenXH2025review_NSR}
Wang, Y. et~al.
\newblock Recent progress in nickelate superconductors.
\newblock {\em National Science Review} 12, nwaf373 (2025).

\bibitem{HYHwang2019DFLi}
Li, D. et~al.
\newblock {Superconductivity in an infinite-layer nickelate}.
\newblock {\em Nature} 572, 624--627 (2019).

\bibitem{DFLi2021}
Li, D. et~al.
\newblock {The discovery and research progress of the nickelate superconductors}.
\newblock {\em Scientia Sinica Physica, Mechanica $\&$ Astronomica} 51, 047405 (2021).

\bibitem{NormanMR2020PRX_Ni112}
Botana, A.~S.  and  Norman, M.~R.
\newblock Similarities and differences between $\text{LaNiO}_2$ and $\text{CaCuO}_2$ and implications for superconductivity.
\newblock {\em Physical Review X} 10, 011024 (2020).

\bibitem{LechermannF2020PRB_Ni112}
F.~Lechermann.
\newblock Late transition metal oxides with infinite-layer structure: nickelates versus cuprates.
\newblock {\em Physical Review B} 101, 081110 (2020).

\bibitem{SakakibaraH2020PRL_Ni112}
Sakakibara, H. et~al. 
\newblock Model construction and a possibility of cuprate-like pairing in a new $d^9$ nickelate superconductor $(\text{Nd, Sr})\text{NiO}_2$.
\newblock {\em Physical Review Letters} 125, 077003 (2020).

\bibitem{WuXX2020PRB_Ni112}
Wu, X. et~al.
\newblock Robust ${d}_{{x}^{2}\ensuremath{-}{y}^{2}}$-wave superconductivity of infinite-layer nickelates.
\newblock {\em Physical Review B} 101, 060504 (2020).

\bibitem{ZhangGM2020PRB_Ni112}
Zhang, G.-M. et~al. 
\newblock Self-doped mott insulator for parent compounds of nickelate superconductors.
\newblock {\em Physical Review B} 101, 020501 (2020).

\bibitem{YPQi2025MXZhang_PRX}
Zhang, M. et~al.
\newblock Superconductivity in trilayer nickelate $\text{La}_4\text{Ni}_3\text{O}_{10}$ under pressure.
\newblock {\em Physical Review X} 15, 021005 (2025).

\bibitem{XHChen2025MZShi_NP}
Shi, M. et~al.
\newblock Pressure-induced superconductivity in hybrid Ruddlesden-Popper $\text{La}_5\text{Ni}_3\text{O}_{11}$ single crystals.
\newblock {\em Nature Physics} 21, 1780–1786 (2025). 

\bibitem{WangM2025HuangC_arXiv}
Huang, C. et~al. 
\newblock Superconductivity in monolayer-trilayer phase of $\text{La}_3\text{Ni}_2\text{O}_{7}$ under high pressure.
\newblock arXiv:2510.12250 (2025).


\bibitem{MitchellJF2020ZhangJJ_NC}
Zhang, J. et~al.
\newblock Intertwined density waves in a metallic nickelate.
\newblock {\em Nature Communications} 11, 6003 (2020).

\bibitem{WangM2024HuangX_CPL}
Huang, X. et~al.
\newblock Signature of superconductivity in pressurized trilayer-nickelate $\text{Pr}_4\text{Ni}_3\text{O}_{10-\delta}$.
\newblock {\em Chin. Phys. Lett.} 41, 127403 (2024).

\bibitem{FengDL2024ChenXY_NC}
Chen, X. et~al.
\newblock Electronic and magnetic excitations in $\text{La}_3\text{Ni}_2\text{O}_{7}$.
\newblock {\em Nature Communications} 15, 9597 (2024).

\bibitem{MukudaH2024_JPSJ}
Kakoi, M. et~al.
\newblock Multiband metallic ground state in multilayered nickelates $\text{La}_3\text{Ni}_2\text{O}_7$ and $\text{La}_4\text{Ni}_3\text{O}_{10}$ probed by $^{139}\text{La}$-NMR at ambient pressure.
\newblock {\em Journal of the Physical Society of Japan} 93, 053702 (2024).

\bibitem{ShuL2024PRL_miuSR}
Chen, K. et~al. 
\newblock Evidence of spin density waves in $\text{La}_3\text{Ni}_2\text{O}_{7-\delta}$.
\newblock {\em Physical Review Letters} 132, 256503 (2024).

\bibitem{ZhuZH2025CP}
Ren, X. et~al. 
\newblock Resolving the electronic ground state of $\text{La}_3\text{Ni}_2\text{O}_{7-\delta}$ films.
\newblock {\em Communications Physics} 8, 52 (2025).

\bibitem{WangNL2025XuSX_PRB}
Xu, S. et~al.
\newblock Origin of the density wave instability in trilayer nickelate $\mathrm{L}{\mathrm{a}}_{4}\mathrm{N}{\mathrm{i}}_{3}{\mathrm{O}}_{10}$ revealed by optical and ultrafast spectroscopy.
\newblock {\em Physical Review B} 111, 075140 (2025).

\bibitem{YanYJ2025LiMZ_PRB}
Li, M. et~al.
\newblock Direct visualization of an incommensurate unidirectional charge density wave in $\mathrm{L}{\mathrm{a}}_{4}\mathrm{N}{\mathrm{i}}_{3}{\mathrm{O}}_{10}$.
\newblock {\em Physical Review B} 112, 045132 (2025).

\bibitem{GuoHJ2025CaoYT}
Cao, Y. et~al. 
\newblock Complex spin-density-wave ordering in La$_4$Ni$_{3}$O$_{10}$.
\newblock arXiv:2503.14128 (2025).

\bibitem{ZhaoJ2025ZhangEK_PRX}
Zhang, E. et~al. 
\newblock Bulk superconductivity in pressurized trilayer nickelate $\text{Pr}_4\text{Ni}_3\text{O}_{10}$ single crystals.
\newblock {\em Physical Review X} 15, 021008 (2025).

\bibitem{GuguchiaZ2025KhasanovR}
Khasanov, R. et~al. 
\newblock Pressure-enhanced splitting of density wave transitions in $\text{La}_3\text{Ni}_2\text{O}_{7-\delta}$.
\newblock {\em Nature Physics} 21, 430--436 (2025).

\bibitem{YangLX2025LiYD_SB}
Li, Y. et~al.
\newblock Distinct ultrafast dynamics of bilayer and trilayer nickelate superconductors regarding the density-wave-like transitions.
\newblock {\em Science Bulletin} 70, 180--186 (2025).

\bibitem{ChenXH2025ShiMZ_NC}
Shi, M. et~al.
\newblock Absence of superconductivity and density-wave transition in ambient-pressure tetragonal $\text{La}_4\text{Ni}_3\text{O}_{10}$.
\newblock {\em Nature Communications} 16, 2887 (2025).

\bibitem{WangNL2025XuSX_NC}
Xu, S. et~al.
\newblock Collapse of density wave and emergence of superconductivity in pressurized-$\text{La}_4\text{Ni}_3\text{O}_{10}$ evidenced by ultrafast spectroscopy.
\newblock {\em Nature Communications} 16, 7039 (2025).

\bibitem{ChenXH2025arXiv_327}
Shi, M. et~al.
\newblock Prerequisite of superconductivity: SDW rather than tetragonal structure in double-layer $\text{La}_3\text{Ni}_2\text{O}_{7-x}$.
\newblock arXiv:2501.14202 (2025).

\bibitem{GuguchiaZ2025arXiv_miuSR}
Khasanov, R. et~al. 
\newblock Identical suppression of spin and charge density wave transitions in $\text{La}_4\text{Ni}_3\text{O}_{10}$ by pressure.
\newblock arXiv:2503.04400 (2025).

\bibitem{DSDessau2017HXLi}
Li, H. et~al.
\newblock {Fermiology and electron dynamics of trilayer nickelate $\text{La}_4\text{Ni}_3\text{O}_{10}$}.
\newblock {\em Nature Communications} 8, 704 (2017).

\bibitem{YangLX2024DuX_arXiv}
Du, X. et~al.
\newblock Correlated electronic structure and density-wave gap in trilayer nickelate $\text{La}_4\text{Ni}_3\text{O}_{10}$.
\newblock arXiv:2405.19853 (2024).

\bibitem{MitchelJ2020ZhangJJ_PRM}
Zhang, J. et~al. 
\newblock High oxygen pressure floating zone growth and crystal structure of the metallic nickelates $R_4\text{Ni}_3\text{O}_{10}$ ($R = \text{La, Pr}$)).
\newblock {\em Physical Review Materials} 4, 083402 (2020).

\bibitem{WangM2024SCPMA_Ni4310}
Li, J. et~al. 
\newblock Structural transition, electric transport, and electronic structures in the compressed trilayer nickelate $\text{La}_4\text{Ni}_3\text{O}_{10}$.
\newblock {\em Science China Physics, Mechanics $\&$ Astronomy} 67, 117403 (2024).

\bibitem{YangJG2024_NC}
Yang, J. et~al. 
\newblock Orbital-dependent electron correlation in double-layer nickelate $\text{La}_3\text{Ni}_2\text{O}_7$.
\newblock {\em Nature Communications} 15, 4373 (2024).

\bibitem{Damascelli2025arXiv}
Au-Yeung, C. C. et~al. 
\newblock Universal electronic structure of multi-layered nickelates via oxygen-centered planar orbitals.
\newblock arXiv:2502.20450 (2025).

\bibitem{XJZhou2023LuoXY}
Luo, X. et~al. 
\newblock Electronic origin of high superconducting critical temperature in trilayer cuprates.
\newblock {\em Nature Physics} 19, 1841–1847 (2023).

\bibitem{QMSi2016Review_NatureRP}
Paschen, S and Si, Q.
\newblock Quantum phases driven by strong correlations.
\newblock {\em Nature Reviews Physics} 3, 9--26 (2021).

\bibitem{SchmalianJ2025Review}
Esterlis, I. and Schmalian, J.
\newblock Quantum critical Eliashberg theory.
\newblock arXiv:2506.11952 (2025).

\bibitem{ZhangGM2023CPL}
Shen, Y. et~al. 
\newblock Effective bi-layer model Hamiltonian and density-matrix renormalization group study for the high-T$_c$ superconductivity in $\text{La}_3\text{Ni}_2\text{O}_7$ under high pressure.
\newblock {\em Chinese Physics Letters} 40, 127401 (2023).

\bibitem{WangQH2023YangQG_PRB}
Yang, Q.-G. et~al. 
\newblock Possible ${s}_{\ifmmode\pm\else\textpm\fi{}}$-wave superconductivity in $\text{La}_3\text{Ni}_2\text{O}_7$.
\newblock {\em Physical Review B} 108, L140505 (2023).

\bibitem{HuJP2025GuYH_PRB}
Gu, Y. et~al. 
\newblock Effective model and pairing tendency in the bilayer Ni-based superconductor $\text{La}_3\text{Ni}_2\text{O}_7$.
\newblock {\em Physical Review B} 111, 174506 (2025).

\bibitem{YaoDX2024LuoZH_NPJ}
Luo, Z. et~al. 
\newblock High-T$_c$ superconductivity in $\text{La}_3\text{Ni}_2\text{O}_7$ based on the bilayer two-orbital t-J model.
\newblock {\em npj Quantum Materials} 9, 61 (2024).

\bibitem{YangF2023LiuYB_PRB}
Liu, Y.-B. et~al. 
\newblock $s_{\pm}$-wave pairing and the destructive role of apical-oxygen deficiencies in $\text{La}_3\text{Ni}_2\text{O}_7$ under pressure.
\newblock {\em Physical Review Letters} 131, 236002 (2023).

\bibitem{WangQH2024YangQG_PRB_4310}
Yang, Q.-G. et~al. 
\newblock Effective model and ${s}_{\ifmmode\pm\else\textpm\fi{}}$-wave superconductivity in trilayer nickelate $\text{La}_4\text{Ni}_3\text{O}_{10}$.
\newblock {\em Physical Review B} 109, L220506 (2024).

\bibitem{KurokiK2024PRB}
Sakakibara, H. et~al. 
\newblock Theoretical analysis on the possibility of superconductivity in the trilayer Ruddlesden-Popper nickelate $\text{La}_4\text{Ni}_3\text{O}_{10}$ under pressure and its experimental examination: comparison with $\text{La}_3\text{Ni}_2\text{O}_7$.
\newblock {\em Physical Review B} 109, 144511 (2024).

\bibitem{YangF2025ZhangM_PRB}
Zhang, M. et~al. 
\newblock Spin-density wave and superconductivity in $\text{La}_4\text{Ni}_3\text{O}_{10}$ under ambient pressure.
\newblock {\em Physical Review B} 111, 144502 (2025).

\bibitem{ZhouT2024_PRB}
Huang, J. and Zhou, T.
\newblock Interlayer pairing-induced partially gapped Fermi surface in trilayer $\text{La}_4\text{Ni}_3\text{O}_{10}$ superconductors.
\newblock {\em Physical Review B} 110, L060506 (2024).

\bibitem{YangF2024ZhangM_PRB}
Zhang, M. et~al. 
\newblock $s_{\pm}$-wave superconductivity in pressurized $\text{La}_4\text{Ni}_3\text{O}_{10}$.
\newblock {\em Physical Review B} 110, L180501 (2024).

\bibitem{WuCJ2025LuC_PRB}
Lu, C. et~al. 
\newblock Superconductivity in $\text{La}_4\text{Ni}_3\text{O}_{10}$ under pressure.
\newblock {\em Physical Review B} 111, 134515 (2025).

\bibitem{LeonovIV2024PRB_4310}
Leonov, I.~V. 
\newblock Electronic structure and magnetic correlations in the trilayer nickelate superconductor $\text{La}_4\text{Ni}_3\text{O}_{10}$ under pressure.
\newblock {\em Physical Review B} 109, 235123 (2024).

\bibitem{XiangT2024FanZ_PRB}
Fan, Z. et~al. 
\newblock Superconductivity in nickelate and cuprate superconductors with strong bilayer coupling.
\newblock {\em Physical Review B} 110,024514 (2024).

\bibitem{ZhangFC2024JiangK_CPL}
Jiang, K. et~al. 
\newblock High-temperature superconductivity in $\text{La}_3\text{Ni}_2\text{O}_{7}$.
\newblock {\em Chinese Physics Letters} 41, 017402 (2024).

\bibitem{ChenHH2025_NC}
Xia, C. et~al.
\newblock Sensitive dependence of pairing symmetry on Ni-e$_g$ crystal field splitting in the nickelate superconductor $\text{La}_3\text{Ni}_2\text{O}_{7}$.
\newblock {\em Nature Communications} 16, 1054 (2025).

\bibitem{WuXX2025_arXiv_327}
Le, C. et~al. 
\newblock Opposite-mirror-parity scattering as the origin of superconductivity in strained bilayer nickelates.
\newblock arXiv:2501.14665 (2025).

\bibitem{XJZhou2008GDLiu}
Liu, G. et~al. 
\newblock {Development of a vacuum ultraviolet laser-based angle-resolved photoemission system with a superhigh energy resolution better than 1 meV}.
\newblock {\em Review of Scientific Instruments} 79, 023105--11 (2008).

\bibitem{XJZhou2018}
Zhou, X. et~al. 
\newblock {New developments in laser-based photoemission spectroscopy and its scientific applications: a key issues review}.
\newblock {\em Reports on Progress in Physics} 81, 062101 (2018).

\bibitem{MiaoTM2025arXiv}
Miao, T. et~al. 
\newblock Expansion of momentum space and full 2$\pi$ solid angle photoelectron collection in laser-based angle-resolved photoemission spectroscopy by applying sample bias.
\newblock arXiv:2511.19064 (2025).

\end{thebibliography}

\vspace{3mm}

\noindent {\bf\large Acknowledgements}\\
\begin{sloppypar}
 This work is supported by the National Key Research and Development Program of China (Grant No.  2021YFA1401800, 2022YFA1604200, 2022YFA1403900, 2022YFA1403901, 2023YFA1406002, 2023YFA1406103, 2023YFA1407300, 2024YFA1408301 and 2024YFA1408100), the National Natural Science Foundation of China (Grant No. 12488201, 12374066, 12374154, 12494593, 12494594, 12425404, 12504170, 12574151, 12447103, 12447101, 11920101005, 11888101, 12047503, 12322405 and 12104450), CAS Superconducting Research Project (Grant No. SCZX-0101), Quantum Science and Technology-National Science and Technology Major Project (Grant No. 2021ZD0301800), Synergetic Extreme Condition User Facility (SECUF), Fundamental and Interdisciplinary Disciplines Breakthrough Plan of the Ministry of Education of China (Grant No. JYB2025XDXM403), the Guangdong Provincial Key Laboratory of Magnetoelectric Physics and Devices (Grant No. 2022B1212010008), Research Center for Magnetoelectric Physics of Guangdong Province (Grant No. 2024B0303390001) and the New Cornerstone Investigator Program.
\end{sloppypar}

\vspace{3mm}

\noindent {\bf\large Author Contributions}\\
 X.J.Z., L.Z. and J.G.Y. proposed and designed the research. J.G.Y., Q.C.X., Y.H.L and Y.Y.X. carried out the ARPES experiment. T.M.M., B.L., N.C., H.C., W.P.Z., M.K.X., S.J.Z, F.F.Z., F.Y., Z.M.W., Q.J.P., H.Q.M., X.T.L., Z.H.Z, G.D.L., Z.Y.X., L.Z. and X.J.Z. contributed to the development and maintenance of Laser-ARPES system. M.W.H. and M.W. contributed to single crystal growth and characterizations. J.Z., J.P.H. and X.X.W. carried out the theoretical analysis. J.G.Y., L.Z. and X.J.Z. analyzed the data. J.G.Y., L.Z. and X.J.Z. wrote the paper. All authors participated in the discussion and comment on the paper.

\vspace{3mm}

\noindent{\bf\large Competing interests}\\
The authors declare no competing interests.

\begin{figure*}[tbp]
\begin{center}
\includegraphics[width=0.9\columnwidth,angle=0]{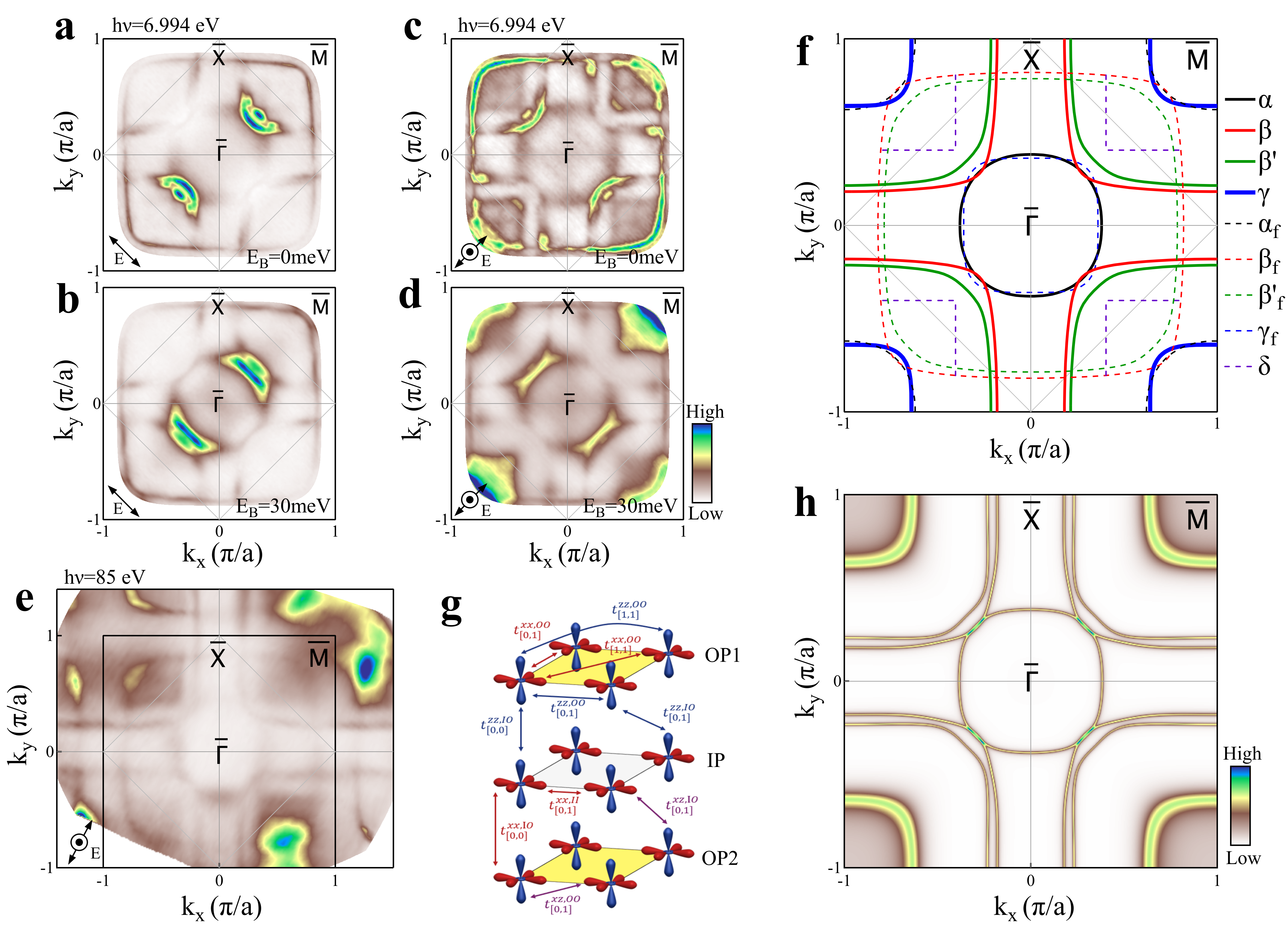}
\end{center}
\caption{\textbf{Measured and simulated Fermi surface of La$_4$Ni$_3$O$_{10}$.} \textbf{a-b} Fermi surface mapping (a) and constant energy contour at the binding energy of 30\,meV (b) measured at 20\,K by using laser-based ARPES with a photon energy of 6.994\,eV under the $s$ polarization geometry where the electric field vector E of the incident light is perpendicular to the photoelectron emission plane; its direction is marked by the double arrow near the bottom-left corner. To increase the momentum coverage, a sample bias voltage of -97 volts was applied during the measurements\cite{MiaoTM2025arXiv}.  \textbf{c-d} Same as (a-b) but measured under $p$ polarization geometry where the electric field vector E is within the photoelectron emission plane. In our case, it consists of both the in-plane component and the out-of-plane component as marked by a double arrow and an out-of-plane arrow near the bottom-left corner. \textbf{e} Fermi surface mapping measured at 20\,K by using synchrotron-based ARPES with a photon energy of 85\,eV.  \textbf{f} Measured Fermi surface of La$_4$Ni$_3$O$_{10}$ obtained from (a), (c), (e) and band structure analysis. \textbf{g} Schematic of trilayer tight-binding model for La$_4$Ni$_3$O$_{10}$. It consists of three Ni-O planes, one inner plane (IP) and two out planes (OP1 and OP2). It involves two orbitals, $d_{x^2-y^2}$ (Red colored) and $d_{z^2}$ (blue colored). The hopping between orbitals is marked by colored double arrows (see details in Supplementary Information). \textbf{h} Simulated Fermi surface from the tight-binding model. The details are described in Supplementary Information.  } 
\end{figure*}

\begin{figure*}[tbp]
\begin{center}
\includegraphics[width=1.0\columnwidth,angle=0]{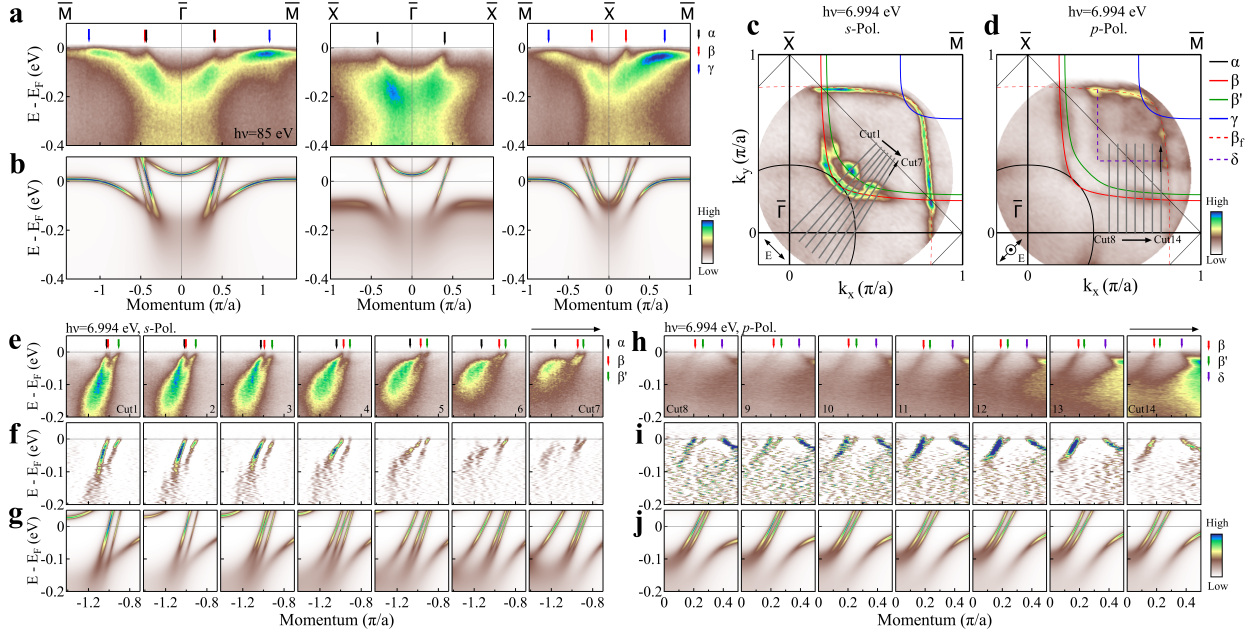}
\end{center}
\caption{\textbf{Measured and simulated band structures of La$_4$Ni$_3$O$_{10}$.} \textbf{a} Band structures along high symmetry directions of $\overline{M}-\overline{\Gamma}-\overline{M}$ (left panel), $\overline{X}-\overline{\Gamma}-\overline{X}$ (middle panel) and $\overline{M}-\overline{X}-\overline{M}$ (right panel) measured at 20\,K by synchrotron-based ARPES with a photon energy of 85\,eV. The observed bands are marked by arrows with different colors. \textbf{b} Simulated band structures along the corresponding high symmetry directions. To illustrate both the occupied and unoccupied electronic states, these simulated data are not multiplied by the Fermi-Dirac distribution function. \textbf{c-d} Measured Fermi surface mappings by laser-based ARPES with 6.994\,eV photon energy in $s$ (c) and $p$ (d) polarization geometries. A sample bias voltage of -30 volts was applied during the measurements. The locations of the momentum cuts are marked by solid grey lines. \textbf{e} Band structures along different momentum cuts crossing M point measured at 20\,K with 6.994\,eV photon energy under $s$ polarization geometry. The location of the momentum cuts is indicated by solid grey lines in (c). \textbf{f} Corresponding MDC second derivative images from (e). \textbf{g} Simulated band structures along the corresponding momentum cuts. \textbf{h} Band structures along different momentum cuts measured at 20\,K with 6.994\,eV photon energy under $p$ polarization geometry. The location of the momentum cuts is indicated by solid grey lines in (d). \textbf{i} Corresponding MDC second derivative images from (h). \textbf{j} Simulated band structures along the corresponding momentum cuts.
}
\end{figure*}

\begin{figure*}[tbp]
	\begin{center}
		\includegraphics[width=1.0\columnwidth,angle=0]{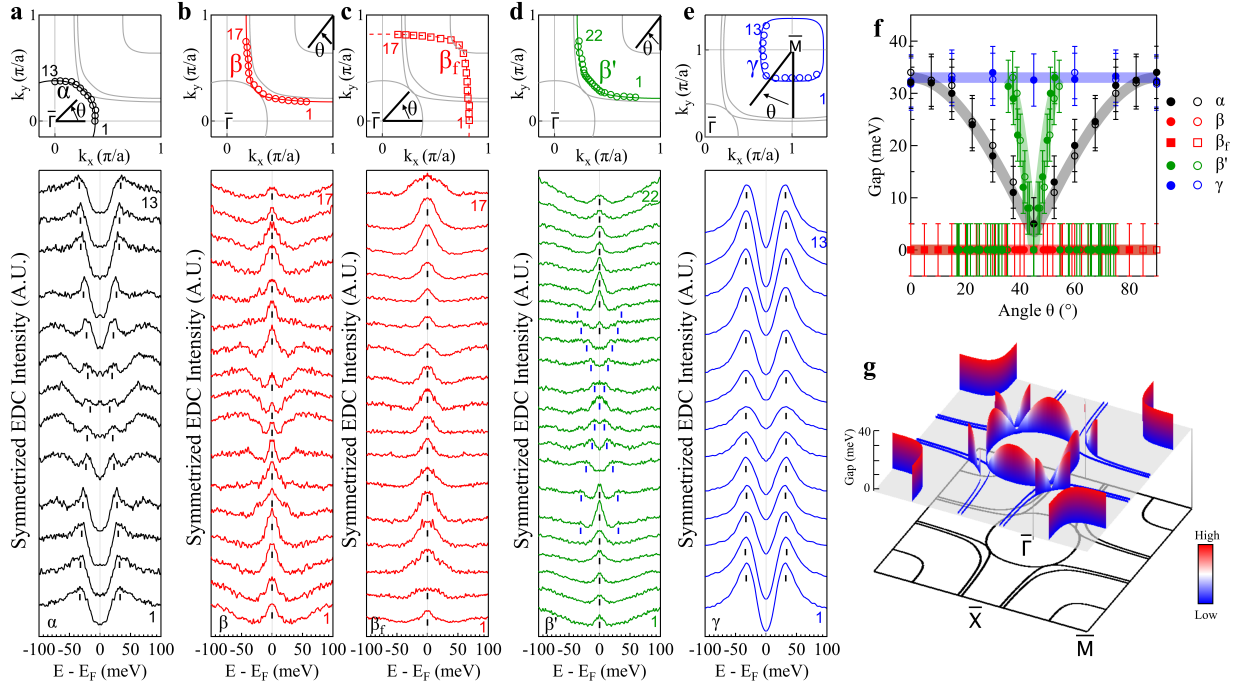}
	\end{center}
	\caption{\textbf{Momentum dependent energy gap along different Fermi surface sheets in La$_4$Ni$_3$O$_{10}$}. \textbf{a} Symmetrized EDCs along the $\alpha$ Fermi surface measured at 20\,K by laser-based ARPES with 6.994\,eV photon energy (lower panel). The location of the momentum points is indicated by empty circles in the upper panel. The EDC peak position is marked by ticks. \textbf{b-e} Same as (a) but along the $\beta$ (b), $\beta_f$ (c) and $\beta'$ (d) Fermi surface. \textbf{e} Symmetrized EDCs along the $\gamma$ Fermi surface measured at 20\,K by synchrotron-based ARPES with 85\,eV photon energy (lower panel). The location of the momentum points is indicated by empty circles in the upper panel. \textbf{f} Extracted energy gap as a function of the Fermi surface angle $\theta$ for the $\alpha$ (black circles), $\beta$ (red circles), $\beta_f$ (red squares), $\beta'$ (green circles) and $\gamma$ (blue circles) Fermi surfaces obtained from (a-e). The Fermi surface angle $\theta$ is defined in the upper panels of (a-e). The solid symbols are derived from the symmetrized EDCs shown in lower panels of (a–e) whereas the open symbols are obtained by symmetrization, taking into account the fourfold symmetry. The uncertainties are marked by error bars. \textbf{g} Three-dimensional plot of the energy gap in La$_4$Ni$_3$O$_{10}$. The corresponding Fermi surface is shown at the bottom.
      } 	
\end{figure*}

\begin{figure*}[tbp]
\begin{center}
\includegraphics[width=1.0\columnwidth,angle=0]{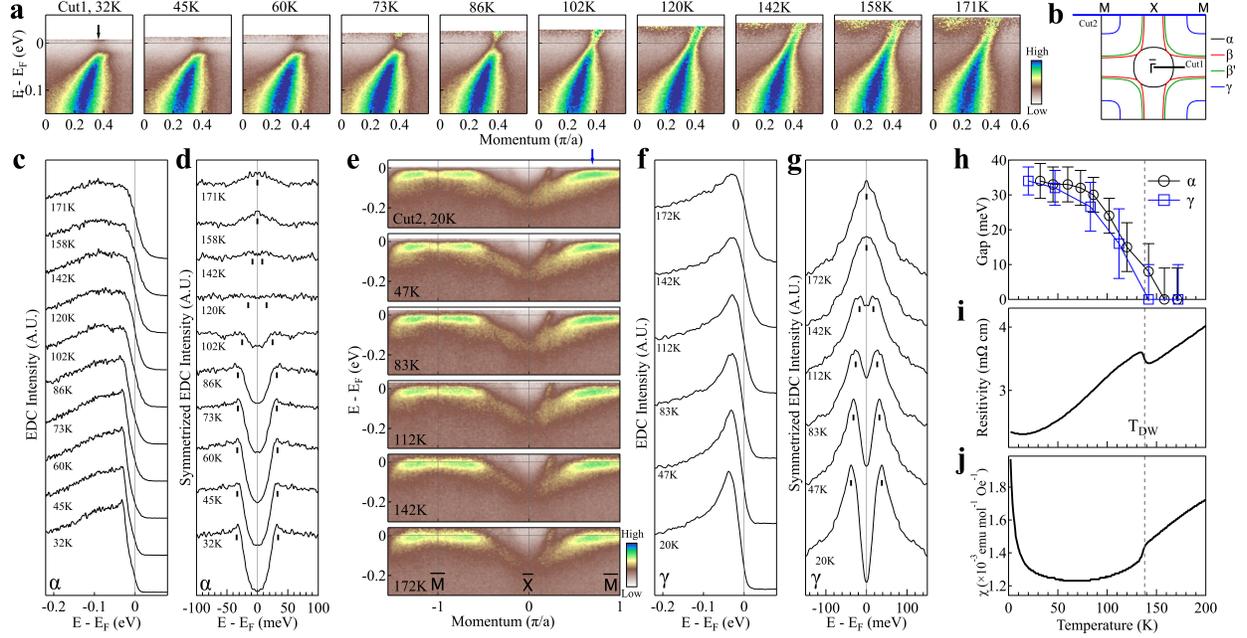}
\end{center}
\caption{\textbf{Temperature evolution of the energy gap on $\alpha$ and $\gamma$ bands}.  \textbf{a} Band structure of the $\alpha$ band measured along the momentum cut (Cut1) at different temperatures. The location of the momentum cut (Cut1) is marked by the black solid line in (b). \textbf{b} Schematic Fermi surface of La$_4$Ni$_3$O$_{10}$. The location of two momentum cuts (Cut1 and Cut2) is marked. \textbf{c} EDCs at the Fermi momentum of the $\alpha$ band measured at different temperatures obtained from (a). The momentum location to extract the EDCs is marked by the black arrow in the left-most panel in (b). \textbf{d}  Corresponding symmetrized EDCs obtained from (c). The black ticks mark the peak position of the symmetrized EDCs. \textbf{e} Band structure of the $\gamma$ band measured along the momentum cut (Cut2) at different temperatures. The location of the momentum cut (Cut2) is marked by the blue solid line in (b).  \textbf{f} EDCs at the Fermi momentum of the $\gamma$ band measured at different temperatures obtained from (e). The momentum location to extract the EDCs is marked by the blue arrow in (e). \textbf{g}  Corresponding symmetrized EDCs obtained from (f). The black ticks mark the peak position of the symmetrized EDCs. \textbf{h} Temperature evolution of the energy gap on the $\alpha$ (black circles) and $\gamma$ (blue squares) bands obtained from (d) and (g), respectively. \textbf{i} Temperature dependent resistivity of La$_4$Ni$_3$O$_{10}$ measured in ab-plane. \textbf{j} Temperature dependent magnetic susceptibility ($\chi$) of La$_4$Ni$_3$O$_{10}$ measured with out-of-plane magnetic field. }

\end{figure*}

\begin{figure*}[tbp]
\begin{center}
\includegraphics[width=1.0\columnwidth,angle=0]{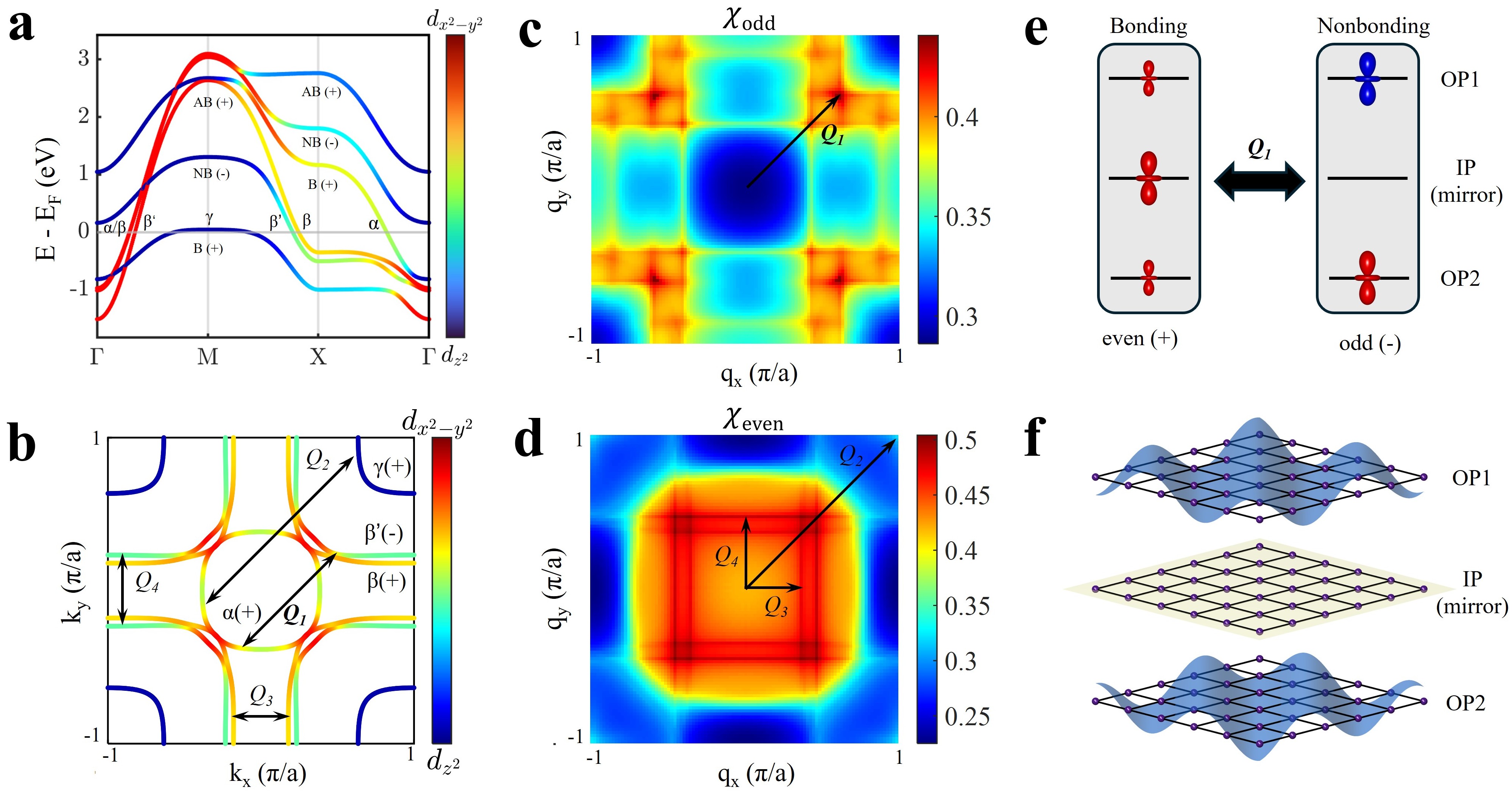}
\end{center}
\caption{\textbf{Mirror-selective scattering driven interlayer antiferromagnetic SDW in La$_4$Ni$_3$O$_{10}$}. \textbf{a} Calculated band structures along high-symmetry directions from the tight-binding model (see Note 1 in Supplementary Information for details and used parameters). These bands are labeled as $\alpha$, $\beta$, $\beta'$ and $\gamma$ with their bonding character (bonding, non-bonding and anti-bonding) and parity (even (+) and odd (-)) marked. The parity is determined by considering the symmetry of the wavefunction of the bands in the three Ni-O layers with respect to the inner mirror plane (see Supplementary Information for details). \textbf{b} Orbital-projected Fermi surface calculated using the tight-binding model. The color scale represents the relative contribution of the $d_{z^2}$ and $d_{x^2-y^2}$ orbitals. The parity of each Fermi surface sheet is labeled as $+$ (even) or $-$ (odd).  The arrowed lines denote the nesting vectors. \textbf{c} Calculated odd-channel susceptibility ($\chi_{odd}$) from the tight-binding model (see Supplementary Information for details). The scattering vector $Q_1\approx(0.62\pi, 0.62\pi)$ is indicated by the arrowed line. \textbf{d} Calculated even-channel susceptibility ($\chi_{even}$). The scattering vectors $Q_2$, $Q_3$ and $Q_4$ are indicated by the arrowed lines. \textbf{e} Schematic of real-space wavefunctions of the bonding and nonbonding states of the $d_{z^2}$ orbitals. Horizontal lines denote the three Ni-O layers. With respect to the inner mirror plane, the wave function of the bonding state exhibit even (+) parity while that of the non-bonding state exhibits odd (-) parity. The $Q_1$ nesting occurs between these two states. \textbf{f} Schematic illustration of the interlayer antiferromagnetic SDW with wave vector $Q_1\approx(0.62\pi, 0.62\pi)$ in La$_4$Ni$_3$O$_{10}$. The blue shapes represent the spin density modulation on the Ni-O layers which exhibits a $\pi$ phase shifts between OP1 and OP2. 
}
\end{figure*}

\end{document}